\documentclass[aps,prd,preprintnumbers,showpacs,showkeys,nofootinbib,superscriptaddress,fleqn,floatfix,tightenlines,10pt]{revtex4-2} 
\usepackage{amsmath,amsfonts,amssymb,amscd,amsxtra,amsthm}
\usepackage{dsfont}
\usepackage{graphicx}  
\usepackage{epstopdf}
\usepackage{dcolumn}  
\usepackage{bm}          
\usepackage{slashed}
\usepackage[utf8]{inputenc}
\usepackage{hyperref}
\usepackage{booktabs}
\usepackage[normalem]{ulem} 
\usepackage[dvipsnames]{xcolor} 
\usepackage{enumitem}
\usepackage{array}
\usepackage{slashed}
\usepackage{tikz}
\usepackage{float}
\usepackage{multirow}
\usepackage{booktabs} 
\usepackage{slashed}

\renewcommand\sout{\bgroup \color{red} \ULdepth=-.5ex \ULset}

\begin{document}
\preprint{INHA-NTG-04/2026}
\title{Probing hidden-charm pentaquarks from the $\pi N\rightarrow
  J/\psi N$ reaction} 
\author{Samson Clymton}
\email[E-mail: ]{samson.clymton@apctp.org}
\affiliation{Asia Pacific Center for Theoretical Physics (APCTP),
  Pohang, Gyeongbuk 37673, Republic of Korea} 

\author{Sun-Young Ryu}
\email[E-mail: ]{syryu@rcnp.osaka-u.ac.jp} 
\affiliation{Research Center for Nuclear Physics, Osaka University, Ibaraki  567-0047,  Japan} 

\author{Jung-Keun Ahn}
\email[E-mail: ]{ahnjk@korea.ac.kr}
\affiliation{Department of Physics, Korea University,
  Seoul , Republic of Korea}

\author{Hyun-Chul Kim}
\email[E-mail: ]{hchkim@inha.ac.kr}
\affiliation{Department of Physics, Inha University,
  Incheon 22212, Republic of Korea}
\affiliation{Institute of Quantum Science, Inha University, Incheon 22212,
  Republic of Korea}  
\affiliation{School of Physics, Korea Institute for Advanced Study 
  (KIAS), Seoul 02455, Republic of Korea}
\date{\today}

\begin{abstract}
We investigate the dynamical generation of hidden-charm pentaquark
resonances in the $\pi N \to J/\psi N$ reaction utilizing an off-shell
coupled-channel formalism. Motivated by the absence of pentaquark
signals in $J/\psi$ photoproduction, we evaluate rescattering effects
with two-body kernel amplitudes constructed from effective Lagrangians
that explicitly incorporate $t$-channel meson and $u$-channel baryon
exchanges. We demonstrate that the $u$-channel $\Lambda_c$ exchange,
of which an analogous contribution is absent in the photoproduction
kernel, greatly enhances the rescattering contributions through the
$\bar{D}^{(*)}\Sigma_c$ intermediate states. Consequently, the
$\bar{D}^{(*)}\Sigma_c^{(*)}$ channels yield contributions of
comparable magnitude to the $\bar{D}^{(*)}\Lambda_c$ channels,
directly leading to prominent pentaquark signals. The partial-wave
analysis reveals that the $P_{c\bar{c}}(4312)$ and
$P_{c\bar{c}}(4457)$ states emerge as clear peak structures with
$J^P=1/2^-$ and $3/2^-$, respectively. In contrast, the
$P_{c\bar{c}}(4380)$ and $J^P=5/2^-$ states are strongly suppressed
because the $\Lambda_c$ exchange does not provide the required tensor
interactions. The result for the total cross section reaches the
microbarn level at the peak positions. 
\end{abstract}

\maketitle

\section{Introduction}
The LHCb Collaboration announced the existence of four hidden-charm
pentaquark states~\cite{LHCb:2015yax, LHCb:2019kea, LHCb:2021chn}: 
\begin{align*}
  & P_{c\bar{c}}(4312):\, m_{P_{c\bar{c}}} =
    (4311.9_{-0.9}^{+7.0})~\mathrm{MeV}/c^2,\;\;
    \Gamma_{P_{c\bar{c}}\to J/\psi\, p} = (10\pm 5)~\mathrm{MeV}, \cr
  & P_{c\bar{c}}(4380):\, m_{P_{c\bar{c}}} =
    (4380\pm 30)~\mathrm{MeV}/c^2,\;\;\;
    \Gamma_{P_{c\bar{c}}\to J/\psi\, p} = (210\pm 90)~\mathrm{MeV},
    \cr 
  & P_{c\bar{c}}(4440):\, m_{P_{c\bar{c}}} =
    (4440_{-5}^{+4})~\mathrm{MeV}/c^2,\;\;\;
    \Gamma_{P_{c\bar{c}}\to J/\psi\, p} =
    (21_{-11}^{+10})~\mathrm{MeV},\cr 
  & P_{c\bar{c}}(4457):\, m_{P_{c\bar{c}}} =
    (4457.3_{-1.8}^{+4.0})~\mathrm{MeV}/c^2,\;\;
    \Gamma_{P_{c\bar{c}}\to J/\psi\, p} =
    (6.4_{-2.8}^{+6.0})~\mathrm{MeV}. 
\end{align*}
These findings have stimulated extensive theoretical research into
heavy pentaquark states (see recent reviews~\cite{Esposito:2016noz,
  Chen:2016spr, Guo:2017jvc, Meng:2022ozq, Chen:2022asf,
  Huang:2023jec, Garcilazo:2025wkt} and references therein). In
contrast, the GlueX Collaboration conducted $J/\psi$ photoproduction
experiments to search for these hidden-charm pentaquark states but
found no signals of the $P_{c\bar{c}}$
structures~\cite{GlueX:2019mkq}. Consequently, since these states have
yet to be independently confirmed by other experiments, the Particle
Data Group (PDG) classifies them as ``one-star ($*$)''
baryons~\cite{PDG}.  

Meanwhile, the LHCb Collaboration also observed two hidden-charm
pentaquark states with strangeness $S=-1$: the $P_{c\bar{c}s}(4338)$
with $m_{P_{c\bar{c}s}}=(4338.2\pm 0.8)~\mathrm{MeV}/c^2$ and
$\Gamma_{P_{c\bar{c}s}\to J/\psi\, \Lambda}=(7.0\pm
1.8)~\mathrm{MeV}$~\cite{LHCb:2022ogu}, and the $P_{c\bar{c}s}(4459)$
with $m_{P_{c\bar{c}s}}=(4458.8_{-3.1}^{+6.0})~\mathrm{MeV}/c^2$ and
an unknown width~\cite{LHCb:2020jpq}. Subsequently, the Belle
Collaboration reported evidence supporting the existence of the
$P_{c\bar{c}s}(4459)$: $M_{P_{c\bar{c}s}} = (4471.7 \pm 4.8 \pm
0.6)~\mathrm{MeV}/c^2$ and $\Gamma = (21.9 \pm 13.1 \pm
2.7)~\mathrm{MeV}$~\cite{Belle:2025pey}. However, the measured mass is
approximately $13~\mathrm{MeV}/c^2$ higher than the LHCb value. This
discrepancy leaves it unclear whether the $P_{c\bar{c}s}(4472)$
observed by Belle is identical to the
$P_{c\bar{c}s}(4459)$. Theoretical work by Clymton et
al.~\cite{Clymton:2025hez} suggests that the $P_{c\bar{c}s}(4472)$
should be treated as a distinct state. This situation reinforces the 
critical need for independent experimental confirmation to accurately
determine the properties of newly observed particles. Therefore,
verifying the four $P_{c\bar{c}}$ hidden-charm pentaquarks through
alternative experiments is essential. 

Because the Japan Proton Accelerator Research Complex (J-PARC)
provides pion beams with momenta up to $20~\mathrm{GeV}/c$, it serves
as an ideal facility for experiments involving charmed hadron
production~\cite{Kim:2014qha, Kim:2015ita}. Recently, Ryu et
al. proposed measuring the $\pi N \to J/\psi N$ reaction at
J-PARC~\cite{Ryu:2025}. This process requires a threshold momentum of 
$8.2~\mathrm{GeV}/c$, well within the the capabilities of the
facility. Previous theoretical studies have also suggested 
searching for hidden-charm pentaquarks in this
reaction~\cite{Lu:2015fva, Kim:2016cxr, Wang:2019dsi}, incorporating
these states as explicit $s$-channel pole diagrams alongside their
$u$-channel counterparts. Interestingly, quantitative estimates vary:
L{\"u}~et al.~\cite{Lu:2015fva} predicted a peak structure magnitude
of approximately $1~\mu\mathrm{b}$, whereas Kim et
al.~\cite{Kim:2016cxr}---guided by upper limits from earlier
experimental data~\cite{Jenkins:1977xb}---calculated a total cross
section of about $1~\mathrm{nb}$ near the pentaquark masses.
However, translating the implications of the GlueX data to the $\pi
N\to J/\psi N$ reaction~\cite{Clymton:2026} yields an estimated cross
section on the order of microbarns ($\mu\mathrm{b}$) for these
resonances, a point we will discuss in detail later. 

In the present work, we investigate the emergence of hidden-charm
pentaquark resonances in the $\pi N \to J/\psi N$ reaction utilizing
an off-shell coupled-channel formalism~\cite{Clymton:2024fbf}. This
approach has been successfully applied to describe the dynamical
generation of axial-vector mesons, including the $a_1$, $b_1$, and
$h_1$ states~\cite{Clymton:2022jmv, Clymton:2023txd,
  Clymton:2024pql}. The model was then extended to investigate the
$D_{s0}^*(2317)$ and $B_{s0}^*$ mesons~\cite{Kim:2023htt}, as well as
to explore doubly charmed and hidden-charm tetraquark
states~\cite{Kim:2025ado, Kim:2026eas}. Furthermore, this theoretical
framework has proven highly useful in revealing how hidden-charm
pentaquarks with strangeness $S=-1,\,-2$, and $-3$ are dynamically
generated~\cite{Clymton:2025hez, Clymton:2025zer, Clymton:2025dzt}. To
strictly focus on the dynamical generation of these resonances, we
have intentionally excluded explicit $s$-channel pole diagrams from
the kernel amplitudes.

In a previous study~\cite{Clymton:2024fbf} on the production mechanism
of hidden-charm pentaquarks, it was demonstrated that six such states
are dynamically generated, four of which correspond to the findings
reported by the LHCb Collaboration. Based on an analysis of their
coupling strengths to relevant channels, these four states can be
interpreted as $\bar{D}\Sigma_c(1/2^-)$, $\bar{D}^*\Sigma_c(1/2^-)$,
$\bar{D}\Sigma_c^*(3/2^-)$, and $\bar{D}^*\Sigma_c(3/2^-)$ molecular
configurations, respectively, arranged in ascending order of their
masses. By analyzing the coupled amplitudes in the complex energy
plane, the pole positions of these states clearly emerge on the second
Riemann sheet, confirming their nature as well-defined resonances. In
addition to these negative-parity molecular states, corresponding
positive-parity states were also identified. Thus, the coupled transition
amplitudes derived in Ref.~\cite{Clymton:2024fbf} can be utilized as
input to investigate the $\gamma N\to J/\psi N$ and $\pi N\to J/\psi
N$ reactions. It has already been shown that incorporating these
amplitudes into the rescattering equation for $J/\psi$ photoproduction
off the nucleon suppresses the pentaquark signals, yielding results
consistent with the GlueX data. Therefore, it is of significant
interest to examine the $\pi N\to J/\psi N$ process by including these
amplitudes as rescattering effects. In this work, we will demonstrate
that two hidden-charm pentaquark states indeed manifest as resonances
within this reaction.   

The present paper is organized as follows: In Sec.~II, we formulate
the $\pi N\to J/\psi N$ reaction within the off-shell coupled-channel
framework, treating the transition amplitudes developed in
Ref.~\cite{Clymton:2024fbf} as rescattering effects. We also construct
the kernel amplitudes for this reaction based on an effective
Lagrangian, explicitly incorporating both meson ($t$-channel) and
baryon ($u$-channel) exchanges. Subsequently, we explain the numerical
method for solving the coupled rescattering equation. Section~III
presents the numerical results for the $\pi N \to J/\psi N$
reaction. Here, we analyze the total cross sections for the open-charm
processes, specifically $\pi N\to \bar{D} (\bar{D}^*) Y_c (Y_c^*)$,
where $Y_c(Y_c^*)$ denotes singly-charmed heavy baryons without
strangeness. Furthermore, we evaluate the partial-wave cross sections,
which exhibit clear signals for two hidden-charm pentaquark
states. Finally, Section~IV is devoted to a discussion of our findings
and their implications for future experiments. 
\section{Formalism}
\label{sec:2}
\begin{figure*}[htp] 
\centering
\includegraphics[scale=1.2]{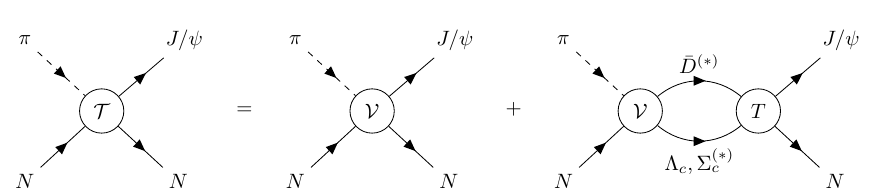}
\caption{Rescattering equations for the $\pi N \to J/\psi N$
  reaction. $\bar{D}^{(*)}$ and $\Sigma_c^{(*)}$ represent generically
$\bar{D}$ ($\bar{D}^*$) mesons and $\Sigma_c$ ($\Sigma_c^*$) baryons.} 
\label{fig:1}
\end{figure*}

Since we focus on the $\pi N \to J/\psi$ reaction in this work, we will
consider the $\pi N$ channel seprately from the $J/\psi N$ and
open-charm meson-baryon channels. Since its threshold is located far
from the charm sector, the coupled-channel effects from elastic $\pi
N$ scattering can safely be ignored. Thus, the transition amplitude
for $\pi N\to J/\psi N$ is expressed in terms of the following
rescattering equation: 
\begin{align}
  \mathcal{T}_{J/\psi N,\pi N} (\bm{p}',\bm{p}) =\, \mathcal{V}_{J/\psi N,\pi N}
  (\bm{p}',\bm{p}) 
  +\frac{1}{(2\pi)^3}\sum_k\int \frac{d^3q}{2E_{k1}(\bm{q})E_{k2}
  (\bm{q})} \mathcal{V}_{k,\pi N}(\bm{q},\bm{p})\frac{E_k
  (\bm{q})}{s-E_k^2(\bm{q})+i\varepsilon} 
  T_{J/\psi N,k}(\bm{p}',\bm{q}),
\label{eq:1}                                                 
\end{align}
which is illustrated schematically in Fig.~\ref{fig:1}.
Here, $\bm{p}$ and $\bm{p}'$ are the relative three-momenta
of the initial and final states,
respectively, while $\bm{q}$
denotes the three-momentum of the intermediate state in the center-of-mass (CM) frame. The variable $s$
represents the square of the total CM energy. 
$E_k$ represents the total on-mass-shell energy of the
intermediate state, $E_k = E_{k1}+E_{k2}$.
The index $k$ runs over
the intermediate two-body open-charm states: $\bar{D}\Lambda_{c}$,
$\bar{D}^{*}\Lambda_{c}$, $\bar{D}\Sigma_{c}$,
$\bar{D}\Sigma_{c}^{*}$, $\bar{D}^{*}\Sigma_{c}$, and
$\bar{D}^{*}\Sigma_{c}^{*}$. 
The $\mathcal{V}_{k,\pi N}$ denotes the two-body Feynman kernel
amplitude for an intermediate state $k$ in the rescattering
equation. When $k=J/\psi  N$, $\mathcal{V}_{J/\psi N,\pi N}$ stands
for the $\pi N\to J/\psi N$ Born diagram. The $T_{J/\psi
  N,k}$ represents the transition amplitude from channel $k$ to the
$J/\psi N$ channel. 

To simplify numerical calculation and provide the partial-wave analysis of the $\pi N \to J/\psi N$ reaction, we
perform the partial-wave decomposition to all amplitudes: 
\begin{align}
\mathcal{T}^{J}_{\lambda'\lambda} (\mathrm{p}',\mathrm{p}) =
\mathcal{V}^{J}_{\lambda'\lambda} (\mathrm{p}',\mathrm{p})+
\frac{1}{(2\pi)^3} \sum_{k,\lambda_k} \int 
\frac{\mathrm{q}^2d\mathrm{q}}{2E_{k1}E_{k2}}
\mathcal{V}^{J}_{\lambda_k\lambda} (\mathrm{q},\mathrm{p})
\frac{E_k}{s-E_k^2+i\varepsilon}
T^{J}_{\lambda'\lambda_k}(\mathrm{p}',\mathrm{q}),
\label{eq:BS-1d}
\end{align}
where the helicities of the final, initial, and intermediate states
are denoted by $\lambda'=\{\lambda'_1,\lambda'_2\}$,
$\lambda=\{\lambda_1,\lambda_2\}$, and
$\lambda_k=\{\lambda_{k1},\lambda_{k2}\}$, respectively. Note that we drop the channel indices to simplify the notation.
The variables $\mathrm{p}'$, $\mathrm{p}$, and $\mathrm{q}$ represent the magnitudes
of the corresponding three-momenta $\bm{p}'$, $\bm{p}$, and $\bm{q}$,
respectively. The partial-wave expansion for the kernel amplitudes
$\mathcal{V}^{J}_{\lambda'\lambda}$ is given by 
\begin{equation}
\mathcal{V}^{J}_{\lambda'\lambda}(\mathrm{p}',\mathrm{p}) =
2\pi \int d(\cos\theta)
d^{J}_{\lambda_1-\lambda_2,\lambda'_1-\lambda'_2}(\theta)
\mathcal{V}_{\lambda'\lambda}(\mathrm{p}',\mathrm{p},\theta),
\label{eq:pwd}
\end{equation}
where $\theta$ is the scattering angle and
$d^{J}_{\lambda\lambda'}(\theta)$ denotes the reduced Wigner $D$
functions. 

Singularities arise from the two-body propagator in the
coupled integral equation in Eq.~\eqref{eq:BS-1d}, which can be
isolated to handle. So, the integral equation is decomposed into the
regularized one and the singular part:  
\begin{align}
  \mathcal{T}^{J}_{\lambda'\lambda} (\mathrm{p}',\mathrm{p}) = 
  \mathcal{V}^{J}_{
  \lambda'\lambda} (\mathrm{p}',\mathrm{p}) + \frac{1}{(2\pi)^3}
  \sum_{k,\lambda_k}\left[\int_0^{\infty}d\mathrm{q}
  \frac{\mathrm{q}E_k}{E_{k1}E_{k2}}\frac{\mathcal{F}(\mathrm{q})
  -\mathcal{F}(\tilde{\mathrm{q}}_k)}{s-E_k^2}+ \frac{1}{2\sqrt{s}}
  \left(\ln\left|\frac{\sqrt{s}-E_k^{\mathrm{thr}}}{\sqrt{s}
  +E_k^{\mathrm{thr}}}\right|-i\pi\right)\mathcal{F}
  (\tilde{\mathrm{q}}_k)\right],
  \label{eq:BS-1d-reg}
\end{align}
where
\begin{align}
  \mathcal{F}(\mathrm{q})=\frac{1}{2}\mathrm{q}\,
  \mathcal{V}^{J}_{\lambda_k\lambda}(\mathrm{q},\mathrm{p})
  T^{J}_{\lambda'\lambda_k}(\mathrm{p}',\mathrm{q}) ,
\end{align}
and $\tilde{\mathrm{q}}_k$ denotes the momentum $\mathrm{q}$ when
$E_{k1}+E_{k2}=\sqrt{s}$. This regularization procedure is exclusively
implemented when the total energy $\sqrt{s}$ is greater than the 
threshold energy of the $k$-th channel $E_k^{\mathrm{thr}}$.

In Ref.~\cite{Clymton:2024fbf}, the transition
amplitudes $T_{J/\psi
  N,k}$ were initially evaluated using a coupled-channel framework
that included only heavy-meson exchange mechanisms. Subsequent
investigation revealed that heavy-baryon exchange contributions are
significant and cannot be neglected. Accordingly, the present study
improves the calculation of the $J/\psi N$ transition amplitudes by
explicitly incorporating these baryon-exchange diagrams. Because the
$J/\psi N$ elastic scattering is kinematically suppressed and the
transitions to the $J/\psi N$ channel are governed by heavy-particle
exchange, this refinement only marginally affects the resonance 
properties. In particular, the pole positions corresponding to the
resonances are almost intact. Thus, we omit a discussion of the
resonance properties themselves in this work.

\begin{figure}[ht]
\centering
\includegraphics[scale=0.4]{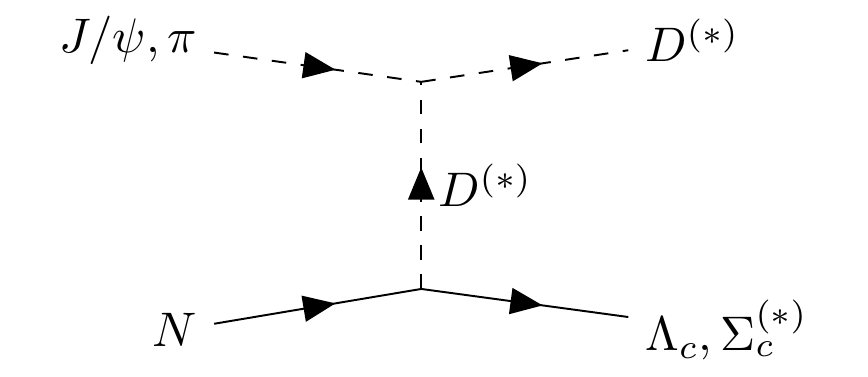}\hspace{0.2cm}
\includegraphics[scale=0.4]{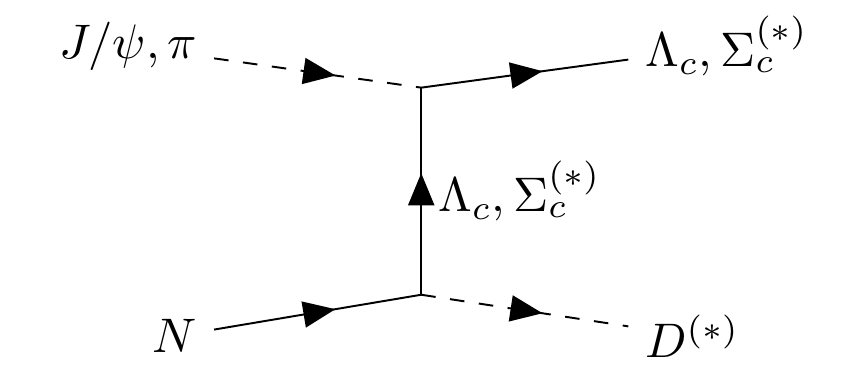}
\caption{Feynman diagrams for the two-body kernel amplitudes
  considered in the present work: $t$-channel 
  meson-exchange diagram (left panel) and $u$-channel baryon-exchange 
  diagram (right panel). } 
\label{fig:2}
\end{figure}
Figure~\ref{fig:2} depicts the Feynman diagrams for the two-body kernel
amplitudes considered in the present work. The left panel of
Fig.~\ref{fig:2} $t$-channel meson-exchange
diagram (left panel) and $u$-channel baryon-exchange  
  diagram (right panel). The exchanged particles considered for each
reaction are summarized in Table~\ref{tab:1}. A kernel amplitude
associated with a given exchange diagram is written as 
\begin{align}
\mathcal{V} = \mathcal{C}_I
  F^{2}(q^{2})\Gamma_{1}(p_{1}^{\prime},p_{2}^{\prime})
  \mathcal{P}(q)\Gamma_{2}(p_{1},p_{2}),  
\end{align}
where $\mathcal{C}_I$ denotes the isospin factor for each exchanged
particle in a given channel, as listed in Table~\ref{tab:1}. The vertex
functions $\Gamma_{1,2}$ are derived from the following effective
Lagrangians with heavy quark spin symmetry:
\begin{align}
\mathcal{L}_{DDJ/\psi} &=
                         -ig_{\psi}M_{D}\sqrt{M_{J}}(J/\psi^{\mu}D^{\dagger}
                         \overleftrightarrow{\partial_{\mu}}D),
                         \cr 
\mathcal{L}_{D^{*}DJ/\psi} &=
                             ig_{\psi}\sqrt{\frac{M_{D}M_{D^{*}}}{M_{J}}}
                             \epsilon^{\mu\nu\alpha\beta}\partial_{\mu}J/
                             \psi_{\nu}(D^{\dagger}
                             \overleftrightarrow{\partial_{\alpha}}D_{\beta}^{*}
                             + D_{\beta}^{*\dagger}
                             \overleftrightarrow{\partial_{\alpha}}D), \cr
\mathcal{L}_{D^{*}D^{*}J/\psi} &= ig_{\psi}M_{D^{*}}
                                 \sqrt{M_{J}}(g^{\mu\nu}g^{\alpha\beta}
                                 -g^{\mu\alpha}g^{\nu\beta}
                                 +g^{\mu\beta}g^{\nu\alpha})(J/\psi_{\mu}
                                 D_{\nu}^{*\dagger}
                                 \overleftrightarrow{\partial_{\alpha}}
                                 D_{\beta}^{*}), \cr
\mathcal{L}_{DN\Lambda_{c}} &= -ig_{I\bar{3}}\sqrt{M_{D}}
                              \bar{N}\gamma_{5}\Lambda_{c}\bar{D}
                              + h.c., \cr
\mathcal{L}_{D^{*}N\Lambda_{c}} &= -ig_{I\bar{3}}
                                  \sqrt{M_{D^{*}}}\bar{N}\gamma^{\mu}
                                  \Lambda_{c}\bar{D}_{\mu}^{*} + h.c., \cr
\mathcal{L}_{DN\Sigma_{c}} &= ig_{I6}\sqrt{\frac{3M_{D}}{2}}
                             \bar{N}\gamma_{5}\bm{\tau}\cdot
                             \bm{\Sigma}_{c}\bar{D} + h.c., \cr
\mathcal{L}_{D^{*}N\Sigma_{c}} &= -ig_{I6}\sqrt{\frac{M_{D^{*}}}{6}}
                                 \bar{N}\gamma^{\mu}\bm{\tau}\cdot
                                 \bm{\Sigma}_{c}\bar{D}_{\mu}^{*} + h.c., \cr
\mathcal{L}_{D^{*}N\Sigma_{c}^{*}} &= ig_{I6}\sqrt{2M_{D^{*}}}
                                     \bar{N}\gamma_{5}\bm{\tau}\cdot
                                     \bm{\Sigma}_{c}^{*\mu}\bar{D}_{\mu}^{*}
                                     + h.c., \cr
\mathcal{L}_{J/\psi\Lambda_{c}\Lambda_{c}} &= -g_{J\bar{3}}
                                             \bar{\Lambda}_{c}\gamma^{\mu}
                                             \Lambda_{c}J/\psi_{\mu}, \cr
\mathcal{L}_{J/\psi\Sigma_{c}\Sigma_{c}} &= -g_{J6}
                                           \bar{\bm{\Sigma}}_{c}\cdot\gamma^{\mu}
                                           \bm{\Sigma}_{c}J/\psi_{\mu}, \cr
\mathcal{L}_{J/\psi\Sigma_{c}\Sigma_{c}^{*}} &=
                                  \frac{2g_{J6}}{\sqrt{3}}\bar{\bm{\Sigma}}_{c}
                                               \cdot\gamma_{5}\bm{\Sigma}_{c}^{*\mu}
                                               J/\psi_{\mu} + h.c., \cr
\mathcal{L}_{J/\psi\Sigma_{c}^{*}\Sigma_{c}^{*}} &=
                       g_{J6}\bar{\bm{\Sigma}}_{c\mu}^{*}
                       \cdot\gamma^{\nu}\bm{\Sigma}_{c}^{*\mu}
                                                   J/\psi_{\nu}.
\label{eq:3}                                                   
\end{align}

\begin{table}[htbp]
  \caption{\label{tab:1}
    Isospin factors ($\mathcal{C}_I$) for the exchange diagrams
    contributing to the transitions from the $J/\psi N$ channel to
    each open-charm meson-baryon channel, together with 
  the corresponding exchanged particles.}  
  \begin{ruledtabular}
  \centering\begin{tabular}{lcr}
   Reactions & Exch. & $\mathcal{C}_I$
   \\\hline
     $J/\psi N\to\bar{D}\Lambda_c$ 
     & $\bar{D}$, $\bar{D}^*$, $\Lambda_c$ & $1$ \\
     $J/\psi N\to\bar{D}^*\Lambda_c$ 
     & $\bar{D}$, $\bar{D}^*$, $\Lambda_c$ & $1$ \\
     $J/\psi N\to\bar{D}\Sigma_c$ 
     & $\bar{D}$, $\bar{D}^*$, $\Sigma_c$ & $\sqrt{3}$ \\
     $J/\psi N\to\bar{D}\Sigma_c^*$ 
     & $\bar{D}^*$, $\Sigma_c$ & $\sqrt{3}$ \\
     $J/\psi N\to\bar{D}^*\Sigma_c$ 
     & $\bar{D}$, $\bar{D}^*$, $\Sigma_c$, $\Sigma_c^*$ & $\sqrt{3}$ \\
     $J/\psi N\to\bar{D}^*\Sigma_c^*$ 
     & $\bar{D}^*$, $\Sigma_c$, $\Sigma_c^*$ & $\sqrt{3}$ \\
  \end{tabular}
    \end{ruledtabular}
\end{table}

Due to the lack of empirical data for the relevant coupling
constants, $SU(4)$ symmetry relations are often used to relate
coupling constants for heavy-to-light hadrons to light ones, which
brings about main uncertainties of the present work.
Fortunately, the GlueX data provides a guideline to determine the
coupling constants for heavy-to-light hadrons.   
We introduce a common prefactor $a$ and multiply all couplings by
it. Thus, the couplings $g_{I\bar{3}}$ and $g_{I6}$, which correspond
respectively to each flavor representation, are then estimated as 
\begin{align}
  g_{I\bar{3}}\sqrt{M_{D}} = -\frac{3\sqrt{3}}{5}ag_{\pi NN}, \quad
  g_{I6}\sqrt{\frac{3M_{D}}{2}} = -\frac{ag_{\pi NN}}{5}.
\end{align}
Similarly, $g_{J\bar{3}}$ and $g_{J6}$ are obtained from
$g_{\rho NN}$ through the $SU(4)$ relation:   
\begin{align}
g_{J\bar{3}} = g_{J6} = \sqrt{2}ag_{\rho NN}.
\end{align}
In the present work, the coupling constants are adopted from the
Nijmegen potential~\cite{Rijken:1998yy,Stoks:1999bz}, namely $g_{\pi
  NN}=13.2$ and $g_{\rho NN}=2.97$. For the mesonic sector, the same
procedure is followed, and the coupling $g_{\psi}$ is estimated as 
\begin{align}
g_{\psi}M_{D}\sqrt{M_{J}} = \frac{\sqrt{2}}{2}ag_{\pi\pi\rho}.
\end{align}
The value $g_{\pi\pi\rho}=5.97$ is taken from our previous
work~\cite{Clymton:2022jmv}. Note that the prefactor $a$ was already
determined to be $a=0.47$ from $J/\psi$ photoproduction off the
nucleon~\cite{Clymton:2026}.   

The propagators for pseudoscalar and vector meson exchanges are given
by 
\begin{align}
\mathcal{P}(q,M) &= \frac{1}{q^{2}-M^{2}}, \cr
\mathcal{P}_{\mu\nu}(q,M) &=
                            \frac{1}{q^{2}-M^{2}}\left(-g_{\mu\nu}+
                            \frac{q_{\mu}q_{\nu}}{M^{2}}\right), 
\end{align}
while those for spin-$1/2$ and spin-$3/2$ baryon exchanges are written
as 
\begin{align}
\mathcal{P}(q,M) &= \frac{\slashed{q}+M}{q^{2}-M^{2}}, \cr
\mathcal{P}_{\mu\nu}(q,M) &= \frac{\slashed{q}+M}{q^{2}-M^{2}}
              \left(-g_{\mu\nu}+\frac{1}{3}\gamma_{\mu}\gamma_{\nu}
               +\frac{1}{3M}(\gamma_{\mu}q_{\nu}-\gamma_{\nu}q_{\mu})
               +\frac{2}{3M^{2}}q_{\mu}q_{\nu}\right),
\end{align}
where $q$ and $M$ denote the four-momentum and mass of the
exchanged particle, respectively.

We introduce a form factor at each vertex, since hadrons have finite
sizes. This form factor is also necessary to ensure the unitarity of
the transition amplitudes. We use the following
form~\cite{Kim:1994ce}:
\begin{align}
F_{n}(q^{2}) =
  \left(\frac{n\Lambda^{2}-M^{2}}{n\Lambda^{2}-q^{2}}\right)^{n},
  \label{eq:9}
\end{align}
where the positive integer value of $n$ is selected such that the
momentum power present in the vertex function $\Gamma$ is tamed.
As $n\to \infty$, the form factor in Eq.~\eqref{eq:9} becomes a
Gaussian type. Note that we need to turn off the 
energy dependence in Eq.~\eqref{eq:9}, which causes unphysical
behavior of a kernel amplitude. 
Although it is difficult to determine the values of the cutoff masses
$\Lambda$ in Eq.~\eqref{eq:9} experimentally or theoretically, we are
able to reduce the uncertainties by considering physical 
properties of the hadrons involved. Considering the fact that 
sizes of heavy hadrons are more compact than those of lighter
ones~\cite{Kim:2018nqf, Kim:2021xpp}, higher cutoff masses are more
suitable for heavy hadrons than for light ones, 
since the cutoff mass is proportional to the inverse of the size of
the corresponding hadron. Thus, we introduce the reduced cutoff mass
as $\Lambda_0 := \Lambda - M$. In Ref.~\cite{Cheng:2004ru}, it was
shown that the cutoff mass can be related to 
$\Lambda_{\mathrm{QCD}}$ via $\Lambda = M + \eta
\Lambda_{\mathrm{QCD}}$, where $\eta$ is of
order unity. This method has been successfully used in previous
works~\cite{Clymton:2022jmv, Clymton:2023txd, Kim:2023htt,
  Clymton:2024pql, Kim:2025ado, Clymton:2024fbf, Clymton:2025hez,
  Clymton:2025zer}. 
In this work, we fix all the values of the reduced cutoff mass to be 
$\Lambda_0 = 600$~MeV. We want to emphasize that we have not performed 
any fitting procedure. 

As mentioned previously, the two-body kernel amplitudes
$\mathcal{V}_{k,\pi N}$ for the initial $\pi N$ and final $k$ states
are constructed from the exchange diagrams depicted in
Fig.~\ref{fig:2}. The isospin factor for each reaction and the
exchanged particles involved are listed in Table~\ref{tab:2}. The
interaction vertices are derived from effective Lagrangians
constructed based on hidden local symmetry, heavy quark spin symmetry,
and $SU(3)$ symmetry. The heavy-baryon and heavy-meson interactions
with light mesons are well established and were employed extensively
in a previous study on hidden-charm
pentaquarks~\cite{Clymton:2024fbf}, where they play a primary
role in generating the pentaquark spectrum. Thus, we use the same form 
of the effective Lagrangians given in Eq.~\eqref{eq:3}, and keep the
same values of the coupling constants.  Once we have constructed the
kernel amplitudes, we can plug them into the rescattering equation of
Eq.~\eqref{eq:1}, and combine them with 
the transition amplitudes for the $k\to J/\psi N$
transitions, where $k=\bar{D}\Lambda_{c}$, 
$\bar{D}^{*}\Lambda_{c}$, $\bar{D}\Sigma_{c}$, $\bar{D}\Sigma_{c}^{*}$,
$\bar{D}^{*}\Sigma_{c}$, $\bar{D}^{*}\Sigma_{c}^{*}$. Having carried
out the partial-wave expansion, we can examine possible hidden-charm
pentaquark states from the $\pi N\to J/\psi N$ reaction.

\begin{table}[htbp]
  \caption{\label{tab:2}
    Isospin factors ($\mathcal{C}_I$) for the
    exchange diagrams contributing to the $\pi N\to
    \bar{D}^{(*)}\Lambda_c$ and $\pi N\to \bar{D}^{(*)}\Sigma_c^{(*)}$
    transitions, together with the type of exchange ($t$- or
    $u$-channel) and the corresponding exchanged particles.}  
  \begin{ruledtabular}
    \renewcommand{\arraystretch}{1.2}
  \centering\begin{tabular}{lccr}
   Reactions & Type & Exchange particles & $\mathcal{C}_I$ \\\hline
     $\pi N\to\bar{D}\Lambda_c$ 
     & $t$ & $\bar{D}^*$  & $-\sqrt{3}$ \\
     & $u$ & $\Sigma_c$, $\Sigma_c^*$  & $-\sqrt{3}$ \\
     $\pi N\to\bar{D}^*\Lambda_c$ 
     & $t$ & $\bar{D}$, $\bar{D}^*$  & $-\sqrt{3}$ \\
     & $u$ & $\Sigma_c$, $\Sigma_c^*$  & $-\sqrt{3}$ \\
     $\pi N\to\bar{D}\Sigma_c$ 
     & $t$ & $\bar{D}^*$ & $1$ \\
     & $u$ & $\Lambda_c$ & $-1$ \\
     & $u$ & $\Sigma_c$  & $2$ \\
     $\pi N\to\bar{D}\Sigma_c^*$ 
     & $t$ & $\bar{D}^*$ & $1$ \\
     & $u$ & $\Lambda_c$ & $-1$ \\
     & $u$ & $\Sigma_c$  & $2$ \\
     $\pi N\to\bar{D}^*\Sigma_c$ 
     & $t$ & $\bar{D}$, $\bar{D}^*$ & $1$ \\
     & $u$ & $\Lambda_c$ & $-1$ \\
     & $u$ & $\Sigma_c$, $\Sigma_c^*$  & $2$ \\
     $\pi N\to\bar{D}^*\Sigma_c^*$ 
     & $t$ & $\bar{D}$, $\bar{D}^*$ & $1$ \\
     & $u$ & $\Lambda_c$ & $-1$ \\
     & $u$ & $\Sigma_c$, $\Sigma_c^*$  & $2$ \\
    \end{tabular}
  \end{ruledtabular}
\end{table}

\section{Results and discussions}
\label{sec:3}
\begin{figure}[ht]
\centering
\includegraphics[scale=0.4]{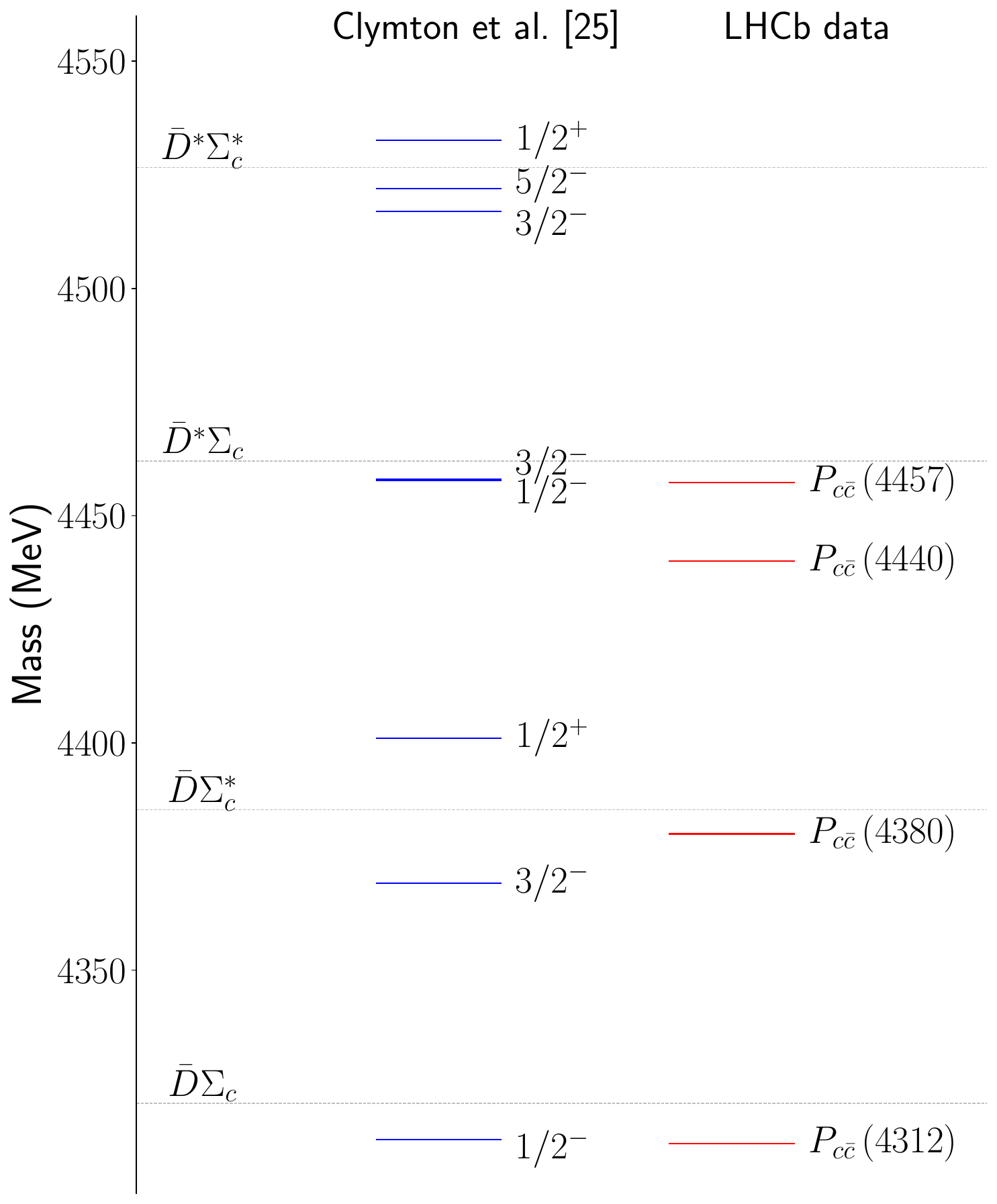}
\caption{Summary of the results from Ref.~\cite{Clymton:2024fbf} in
  comparison with the LHCb data~\cite{LHCb:2015yax, LHCb:2019kea,
    LHCb:2021chn, PDG}.} 
\label{fig:3}
\end{figure}
Before proceeding to discuss the results for the hidden-charm
pentaquark states dynamically generated from the $\pi N \to J/\psi N$
reaction, we recapitulate the main findings of
Ref.~\cite{Clymton:2024fbf}, as shown in Fig.~\ref{fig:3}. Using an
off-shell coupled-channel formalism, six negative-parity hidden-charm
pentaquark states were predicted, as exhibited in the first
column. Four of these six resonances correspond 
to those reported by the LHCb Collaboration, as illustrated in the
second column. Because the lowest state, $P_{c\bar{c}}(4312)$, couples
most strongly to the $\bar{D} \Sigma_c$ channel and lies below the
corresponding threshold, it can be identified as a $\bar{D} \Sigma_c$
molecular state. Similarly, the next state, $P_{c\bar{c}}(4380)$, can
be interpreted as a $\bar{D} \Sigma_c^*$ molecular state. The
remaining two pentaquark states, $P_{c\bar{c}}(4440)$ and
$P_{c\bar{c}}(4457)$, with spins $1/2$ and $3/2$ respectively, are
positioned below the $\bar{D}^*\Sigma_c$ threshold and couple strongly
to the $\bar{D}^*\Sigma_c$ channel. Thus, they can be understood as
$\bar{D}^*\Sigma_c$ molecular states. Furthermore, the $\bar{D}^*
\Lambda_c$, $\bar{D}^* \Sigma_c$, and $\bar{D}^* \Sigma_c^*$ channels
exhibit strong couplings to these pentaquark states, indicating that
coupled-channel effects are essential for describing their
characteristics.

\begin{figure}[ht]
\centering
\includegraphics[scale=0.7]{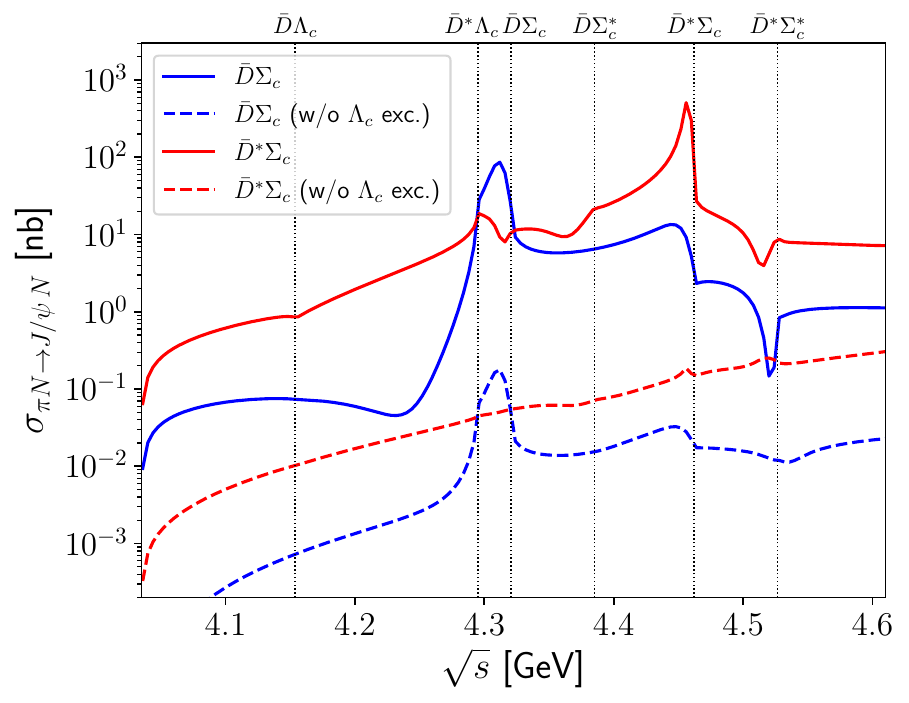}
\caption{Total cross sections for the $\pi N\to J/\psi N$ reaction
  demonstrating rescattering effects. The $\bar{D}\Sigma_c$ and
  $\bar{D}^*\Sigma_c$ contributions are represented by blue and red
  curves, respectively. Dashed lines indicate results excluding the
  $u$-channel $\Lambda_c$ exchange, while solid ones show the results 
  with the $\Lambda_c$ exchange included.}  
\label{fig:4}
\end{figure}
Although the GlueX Collaboration found no signals for hidden-charm
pentaquark states in $J/\psi$ photoproduction off the
proton~\cite{GlueX:2019mkq}, this does not call the existence of these
pentaquarks into question. As described in previous
work~\cite{Clymton:2026}, the $\gamma N$ channel is weakly coupled to
the $\bar{D}^{(*)}\Sigma_c^{(*)}$ channels compared to the
$\bar{D}^{(*)}\Lambda_c^{(*)}$ channels. This weak coupling causes the
suppression of pentaquark signals in $J/\psi$
photoproduction. However, as will be explicitly discussed in this
section, the $\bar{D}^{(*)}\Sigma_c^{(*)}$ channels still play a vital
role in producing hidden-charm pentaquarks in the $\pi N \to J/\psi N$
reaction. The key distinction between the pion-induced reaction and
the photoproduction case lies in the structure of the transition
kernel that drives the $\bar{D}^{(*)}\Sigma_c^{(*)}$ channels: the
$\pi N\to \bar{D}^{(*)}\Sigma_c^{(*)}$ kernel receives a contribution
from the $u$-channel $\Lambda_c$ exchange, which has no counterpart in
the photoproduction kernel. To isolate the role of this contribution,
Figure~\ref{fig:4} compares the rescattering contributions through the
$\bar{D}\Sigma_c$ and $\bar{D}^*\Sigma_c$ intermediate states in the
$\pi N\to J/\psi N$ reaction, evaluated with and without the
$\Lambda_c$ exchange included in the kernel amplitude. As clearly
demonstrated in Fig.~\ref{fig:4}, incorporating the $\Lambda_c$ exchange
greatly amplifies the rescattering effects for both channels. This 
dramatic enhancement indicates that the $\Lambda_c$ exchange serves as
the dominant contribution for populating the $\bar{D}^{(*)}\Sigma_c$
states from the initial $\pi N$ channel. This specific comparison is 
restricted to these two intermediate states because they couple most
strongly to the observed 
$P_{c\bar{c}}$ states. In contrast, the corresponding kernel in
$J/\psi$ photoproduction lacks an analogous contribution; this weak
$\bar{D}^{(*)}\Sigma_c$ coupling precisely suppresses the hidden-charm
pentaquark signals in the $\gamma N\to J/\psi N$ reaction.

\begin{figure}[ht]
\centering
\includegraphics[scale=0.7]{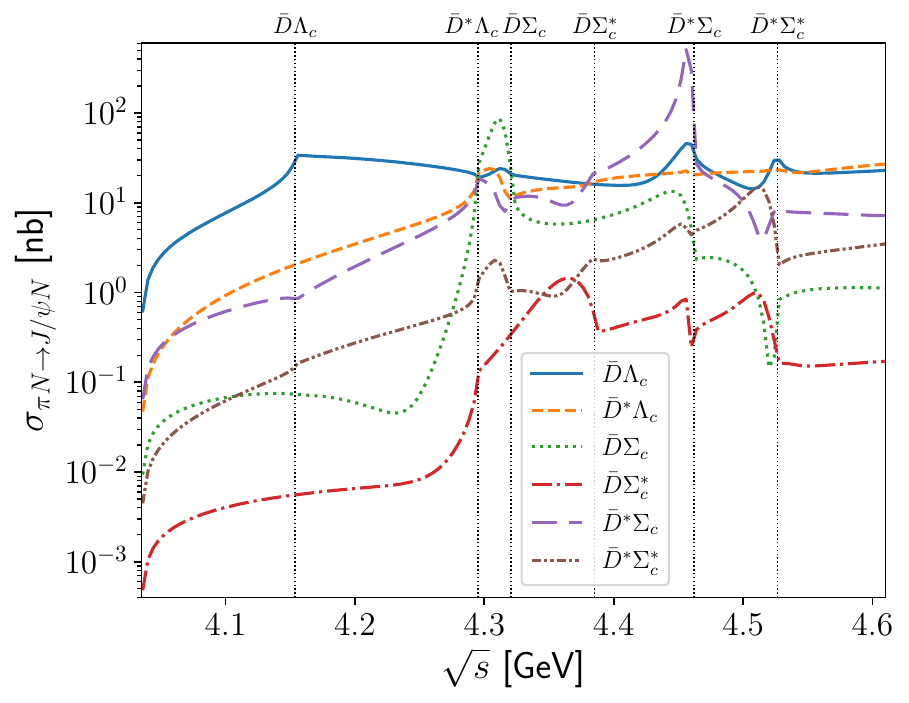}
\caption{Total cross sections of the $\pi N\to J/\psi N$ reaction
  demonstrating rescattering effects from individual
  $\bar{D}^{(*)}\Lambda_c$ and $\bar{D}^{(*)}\Sigma_c^{(*)}$
  intermediate states.}  
\label{fig:5}
\end{figure}
The enhancement observed in Fig.~\ref{fig:4} prompts an
investigation into the relative contributions of the various
intermediate states in the full $\pi N\to J/\psi N$
reaction. Figure~\ref{fig:5} addresses this by presenting the total
cross sections originating from the rescattering effects of each
individual intermediate state, now incorporating the
$\bar{D}^{(*)}\Sigma_c^{*}$ channels alongside the
$\bar{D}^{(*)}\Sigma_c$ channels analyzed in Fig.~\ref{fig:4}. As
reported in Ref.~\cite{Clymton:2026}, the contributions of the
$\bar{D}^{(*)}\Sigma_c$ intermediate states to the photoproduction
process are an order of magnitude smaller than those of the
$\bar{D}^{(*)}\Lambda_c$ channels. In stark contrast, all intermediate
states except for the $\bar{D}\Sigma_c^*$ channel in the pion-induced
reaction yield contributions of comparable magnitude. Because the
$\bar{D}^{(*)}\Sigma_c^{(*)}$ channels dynamically generate the
pentaquark states, their contribution at a 
level comparable to the $\bar{D}^{(*)}\Lambda_c$ channels directly
leads to prominent $P_{c\bar{c}}$ signals in the $\pi N\to J/\psi N$
reaction.

\begin{figure}[ht]
\centering
\includegraphics[scale=0.56]{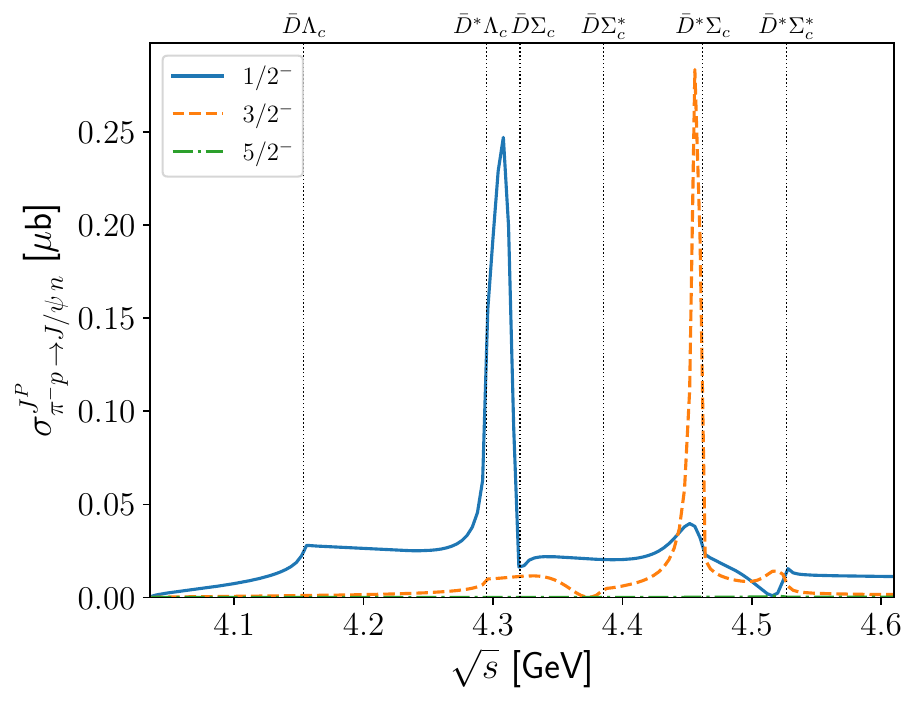}
\includegraphics[scale=0.56]{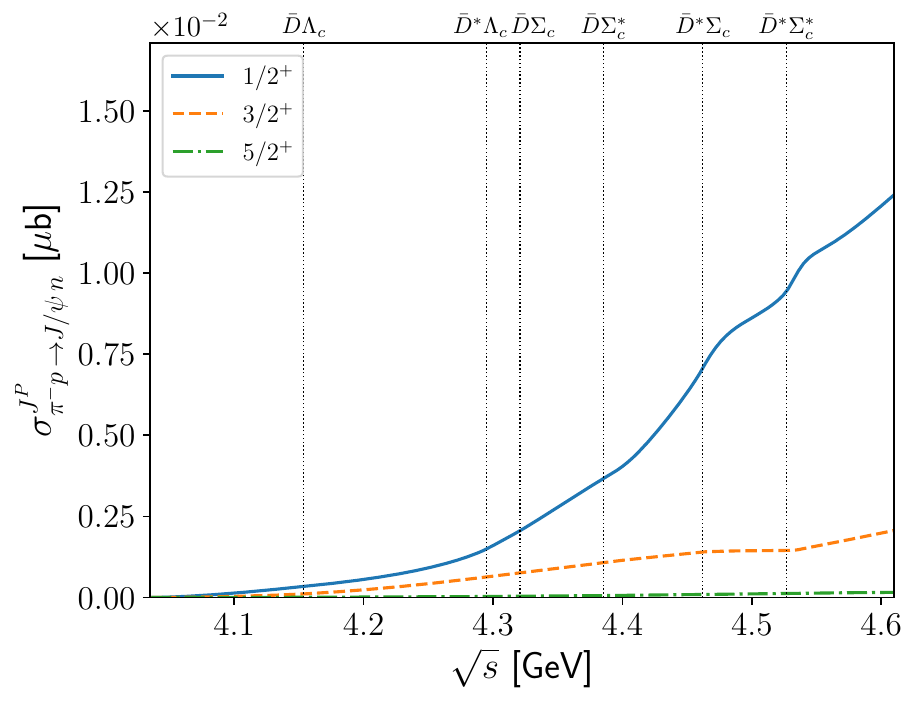}
\caption{Partial-wave cross sections $\sigma^{J^P}$ for the $\pi^- p\to
  J/\psi n$ reaction as functions of the center-of-mass energy 
  $\sqrt{s}$.}  
\label{fig:6}
\end{figure}
In the left panel of Fig.~\ref{fig:6}, we demonstrate the
paratial-wave cross sections with $J^P=1/2^-$, $3/2^-$, and $5/2^-$.
We first clearly observe that the $\sigma^{5/2^-}$ is completely
suppressed. The state with $J^P=5/2^-$ is related to the
$P_{c\bar{c}}(4522)$, which is predicted in
Ref.~\cite{Clymton:2024fbf}. Note that this state is almost solely
coupled to the $\bar{D}^*\Sigma_c^*$ channel. As shown in
Fig.~\ref{fig:5}, the $\pi N\to \bar{D}^* \Sigma_c^*$ transition is
suppressed. Since the $P_{c\bar{c}}(4522)$ has spin $5/2$, the strong
tensor interaction is required to produce it. However, the $\Lambda_c$
exchange in the $u$-channel does not give the tensor interactions, so
that the $\bar{D}^* \Sigma_c^*$ channel must be suppressed.  
Similarly, the signal for the $P_{c\bar{c}}(4380)$ completely
disappears in the partial-wave cross section with $J^P=3/2^-$, since
the  $\pi N\to \bar{D} \Sigma_c^*$ transition is suppressed due to the
lack of the tensor interaction (see Fig.~\ref{fig:5}).
As depicted in the left panel of Fig.~\ref{fig:6}, two
resonances, i.e., $P_{c\bar{c}}(4312)$ and $P_{c\bar{c}}(4457)$,
emerge as clear peaks in the partial-wave cross sections with
$J^P=1/2^-$ and $3/2^-$ respectively. On the other hand, the
$P_{c\bar{c}}(4440)$ with the broad width is dominated by the
$P_{c\bar{c}}(4457)$, so that we cannot observe it in the $\pi N\to
J/\psi N$ reaction. In the right panel of Fig.~\ref{fig:6}, we draw
the results for the partial-wave cross sections with positive parity,
but do not see any signals corresponding to the hidden-charm
pentaquarks with positive parity. 

\begin{figure}[ht]
\centering
\includegraphics[scale=0.7]{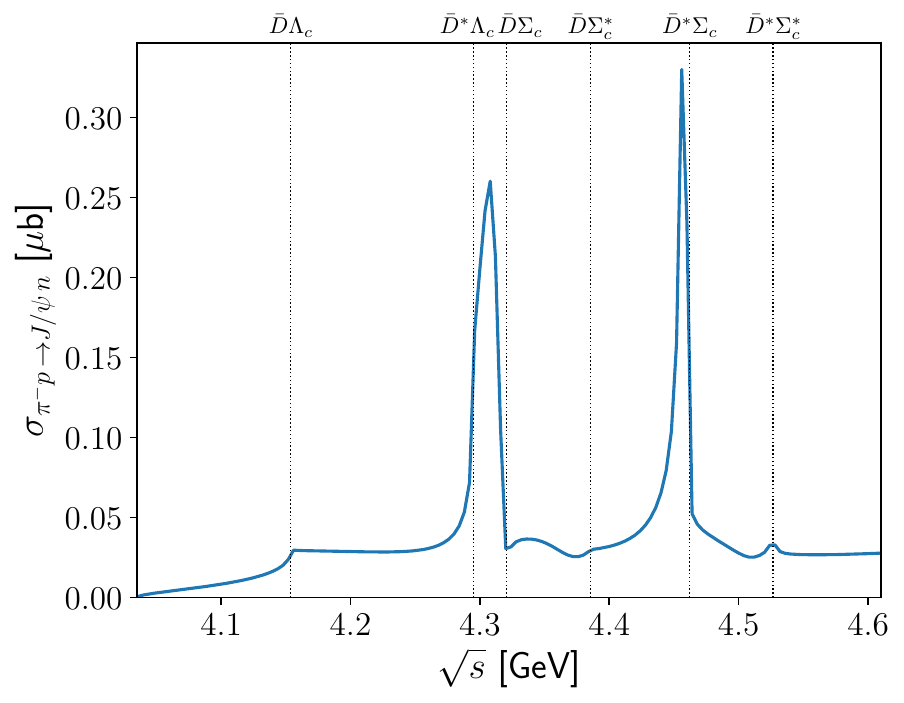}
\caption{Total cross section for the $\pi^- p \to J/\psi n$ reaction
  as a function of the center-of-mass energy $\sqrt{s}$. Prominent
  peak structures corresponding to the $P_{c\bar{c}}(4312)$ and
  $P_{c\bar{c}}(4457)$ states are clearly observed.} 
\label{fig:7}
\end{figure}
To provide concrete predictions and guidance for future experimental
efforts, we calculate the total cross section for the $\pi^- p\to
J/\psi n$ reaction and present the results in Fig.~\ref{fig:7}. Two
pronounced peak structures clearly emerge, corresponding to the
$P_{c\bar{c}}(4312)$ and $P_{c\bar{c}}(4457)$ states. In contrast, the
remaining two pentaquarks reported by the LHCb Collaboration---namely
the $P_{c\bar{c}}(4380)$ and $P_{c\bar{c}}(4440)$---are strongly
suppressed in this process. From an experimental perspective, the
cross section reaches the microbarn level at these peak positions,
dominating the smooth background beneath the resonant structures by
approximately an order of magnitude. Coupled with the clear two-peak
structure, this large signal-to-background ratio establishes the
$\pi^- p\to J/\psi n$ reaction as an effective probe for the 
experimental search and definitive identification of hidden-charm
pentaquark states. 

Last but not least, we want to draw attention to an experimental
proposal of direct relevance to the present theoretical work. The
P111 experiment, situated at the high-energy $\pi20$ beamline of
J-PARC, is designed to deliver high-precision measurements of the
$\pi^-p\to J/\psi\,n$ reaction across a center-of-mass energy range
extending from threshold ($\sqrt{s}=4.04$~GeV) up to $4.54$~GeV.
Reconstruction of the $J/\psi$ meson will proceed through both the
$e^+e^-$ and $\mu^+\mu^-$ dilepton decay channels, taking advantage of
the MARQ (Multipurpose Analyzer to Research Quark-gluon dynamics in
hadrons and hadronic systems) detector, whose electromagnetic
calorimetry and muon identification have been substantially upgraded.
Because this $J/\psi\,n$ final state lies lowest in mass, it offers an
incisive means of probing the internal composition of heavier
resonances that subsequently decay into $J/\psi\,n$. The capability to
determine the $J/\psi$ polarization adds further to the scientific
reach of the measurement. In addition, the sizable leptonic branching
fraction of the $J/\psi$ affords strong rejection of background
originating from multipion processes. Assuming a mean cross section at
the level of a few hundred nanobarns, one anticipates on the order of
$10^4$ events within each $5$~MeV energy bin out to $4.64$~GeV.
Beyond this, the P111 program seeks to uncover $P_{c\bar{c}}$ states by
exploring open-charm reactions such as
$\pi^-p\to\Lambda_c^+\bar{D}^{(\ast)}$. Investigation of these
reactions should expose pronounced features in the center-of-mass
energy spectrum, thereby permitting a sharper extraction of resonance
parameters by way of coupled-channel analyses. Such open-charm
studies may be carried out concurrently with the E50 experiment, which
is devoted to charmed-baryon spectroscopy through the
$p(\pi^-,\bar{D}^\ast)X$ reaction with the same MARQ detector.
\section{Summary and conclusions}
\label{sec:4}
In this work, we investigated the dynamical generation of hidden-charm
pentaquark resonances in the $\pi N \to J/\psi N$ reaction. Motivated
by the absence of $P_{c\bar{c}}$ signals in $J/\psi$ photoproduction,
we employed an off-shell coupled-channel formalism to evaluate the
rescattering effects. We constructed the two-body kernel amplitudes 
using effective Lagrangians, explicitly incorporating $t$-channel
meson and $u$-channel baryon exchanges. 
We demonstrated that the $u$-channel $\Lambda_c$ exchange plays a
crucial role in the pion-induced reaction, providing a mechanism that
has no counterpart in the photoproduction process. This $\Lambda_c$
exchange greatly amplifies the rescattering contributions through the
$\bar{D}^{(*)}\Sigma_c$ intermediate states. Consequently, the
$\bar{D}^{(*)}\Sigma_c^{(*)}$ channels yield contributions of
comparable magnitude to the $\bar{D}^{(*)}\Lambda_c$ channels,
directly leading to prominent pentaquark signals. 
The partial-wave analysis revealed that the $P_{c\bar{c}}(4312)$ and
$P_{c\bar{c}}(4457)$ states emerge as clear peak structures with
$J^P=1/2^-$ and $3/2^-$, respectively. In contrast, the
$P_{c\bar{c}}(4440)$ signal is obscured by the dominant
$P_{c\bar{c}}(4457)$, while the $P_{c\bar{c}}(4380)$ and the
$J^P=5/2^-$ states are strongly suppressed because the $\Lambda_c$
exchange do not provide tensor interactions required to produce them. 
Finally, the calculated total cross section for the $\pi^- p \to
J/\psi n$ reaction reaches the microbarn level at the peak
positions. Coupled with the pronounced two-peak structure, this large
signal-to-background ratio establishes the $\pi^- p \to J/\psi n$
reaction as an effective probe for the experimental search and
definitive identification of hidden-charm pentaquark states. 

\begin{acknowledgments}
The present work was supported by the Young Scientist Training (YST)
Program at the Asia Pacific Center for Theoretical Physics (APCTP)
through the Science and Technology Promotion Fund and Lottery Fund of 
the Korean Government and also by the Korean Local Governments –
Gyeongsangbuk-do Province and Pohang City (SC), the Basic Science
Research Program through the National Research Foundation of Korea
funded by the Korean government (Ministry of Education, Science and
Technology, MEST), Grant-Nos. RS-2025-00513982 (HChK) and  
RS-2024-00436392 (JKA).
\end{acknowledgments}

\bibliography{piN}
\bibliographystyle{apsrev4-2}

\end{document}